# PageRank for ranking authors in co-citation networks

Ying Ding[1], Erjia Yan, Arthur Frazho, James Caverlee

**Abstract**

Google's PageRank has created a new synergy to information retrieval for a better ranking of Web pages. It ranks documents depending on the topology of the graphs and the weights of the nodes. PageRank has significantly advanced the field of information retrieval and keeps Google ahead of competitors in the search engine market. It has been deployed in bibliometrics to evaluate research impact, yet few of these studies focus on the important impact of the damping factor (d) for ranking purposes. This paper studies how varied damping factors in the PageRank algorithm can provide additional insight into the ranking of authors in an author co-citation network. Furthermore, we propose weighted PageRank algorithms. We select 108 most highly cited authors in the information retrieval (IR) area from the 1970s to 2008 to form the author co-citation network. We calculate the ranks of these 108 authors based on PageRank with damping factor ranging from 0.05 to 0.95. In order to test the relationship between these different measures, we compare PageRank and weighted PageRank results with the citation ranking, h-index, and centrality measures. We found that in our author co-citation network, citation rank is highly correlated with PageRank's with different damping factors and also with different PageRank algorithms; citation rank and PageRank are not significantly correlated with centrality measures; and h-index is not significantly correlated with centrality measures.

## 1. Introduction

Google has maintained its continuous success in the search engine market based on its revolutionary and powerful PageRank algorithm. This algorithm manages to bring the most relevant search results to the top of the returned results (Brin & Page, 1998). It assumes web hyperlinks as the trust votes and ranks search results based on these links interlinking them. PageRank creates a new synergy to information retrieval for the better ranking of the Web. It is query independent and content free. It is simple, reliable, and scalable. PageRank spurred a migration in the research world from traditional keyword-based ranking to link-based ranking, further inspiring a variety of new ranking algorithms proposed to improve PageRank (Langville & Meyer, 2006).

PageRank is not a completely new idea. There is a long history of citation research originated from the 1940s which judges the quality of a paper based on citations. Before the advent of the web, printed journals, magazines, or conference proceedings were the main publication channels for academic scholars. Citations, functioning as hyperlinks, interlink scholarly publications and form scholarly graphs. Citation analysis, especially co-citation analysis, constructs an innovative way to analyze and rank documents, authors, or journals (Small, 1973; White & McCain, 1998). In the pioneering paper "The anatomy of a large-scale hypertextual Web search engine," Google founders Sergey Brin and Lawrence Page (Brin & Page, 1998, p. 109), stress the important link between PageRank and citation analysis: "*Academic citation literature has been applied to the web, largely by counting citations or backlinks to a given page. This gives some approximation of a page's importance or quality. PageRank extends this idea by not counting links from all pages equally, and by normalizing by the number of links on a page.*"

---

[1] corresponding author

In bibliometrics, the number of citations is an indicator used to measure the impact of scientific publications. There are some shortcomings concerning this measurement, however, as it does not count the importance of the citing papers: a citation from an obscure paper has the same weight as a citation from a ground-breaking, highly cited work (Maslov & Redner, 2008). This drawback can be alleviated by the PageRank algorithm, since it gives higher weights to the publications that are highly cited (e.g., publications have more inlinks) and also to papers cited by a few highly cited papers (e.g., publications are linked by a few important papers). PageRank is therefore chosen as a complementary method to citation analysis, which allows us to identify publications referenced by highly cited articles.

Two features of the original PageRank algorithm motivate the work in this paper. First, the damping factor, which represents the probability that a random surfer randomly jumps from one page to another page, plays a fundamental role in the PageRank algorithm. It allows the use of a fix point algorithm to compute the vector corresponding to the PageRank. In PageRank, the damping factor is set to be 0.85, meaning an 85% chance that a random surfer will follow the links provided by the present page. This leaves a 15% chance that a random surfer jumps to a completely new page which has no links from the previous surfed pages.

Second, the original PageRank algorithm evenly distributes the PageRank score of one node among its outbound links. These assigned values for outbound links of one node are in turn used to calculate the PageRank value of other nodes which this node points to. So the process is calculated iteratively until it converges. Even though Google's PageRank algorithm has proven to be very successful, this evenly distribution of weights to the outbound links does not necessarily reflect the actual circumstances of the real world, as not all outbound links should have the same importance. Differences do exist in the quality of web pages, researchers, and journals. Most of these domains reflect the power-law distribution phenomena, which indicates that few nodes are truly important and the majority of nodes are merely average. Hence, adding weights to the PageRank algorithm to reflect this observation has begun to attract some research interest recently (Bollen, Rodriquez & Van de Sompel, 2007).

In this paper, we use the PageRank algorithm to rank authors based on the co-citation network, exploring the correlation between various damping factors in this algorithm. We also propose weighted PageRank algorithms, and test these based on the same co-citation network, comparing the results with other traditional bibliometric and social network measures. This research places special focus on whether different values of the damping factor in PageRank algorithm can affect author ranks, and whether the proposed weighted PageRank can produce better ranking results. These PageRank results will be compared with traditional citation analysis (e.g., citation ranks, h-index) and social network analysis (e.g., centrality measures) to verify the correlation among these different ranking methods. This paper is organized as follows: Section 1 introduces the topic; Section 2 provides a review of the PageRank algorithm; Section 3 identifies some related works; Section 4 explains the methods used in this paper; Section 5 discusses the results; Section 6 compares them with other related methods in bibliometrics and social network analysis; and Section 7 concludes the research and identifies future works.

## 2. A review of the PageRank algorithm

In this section, we present a brief review of the PageRank algorithm. To begin, we say that $x$ is a *Markov vector* in $\mathbb{R}^v$ if $x$ is a column vector of length $v$ and the sum of all the components of $x$ equals one.

Moreover, *T* is a *Markov matrix* on $\mathbb{R}^v$ if *T* is a $v \times v$ matrix whose entries are all positive and each column sums to one. In other words, *T* is a Markov matrix if and only if all the columns of *T* are Markov vectors.

The PageRank algorithm can be viewed as a state space system of the form
$$x(n+1) = dTx(n) + b \qquad (1)$$
The state *x(n)* is a Markov vector of length *v*, and *T* is a Markov matrix on $\mathbb{R}^v$, while *d*, the so-called *damping factor*, is a scalar in [0, 1). Furthermore, *b* is a vector of length *v* consisting of all positive numbers which sum to 1- *d*. The initial condition *x*(0) is a Markov vector in $\mathbb{R}^v$. This guarantees that the state *x*(*n*) is also a Markov vector. In PageRank the components $T_{jk}$ of *T* represent the probability that a random surfer will move from webpage *k* to webpage *j* following one of the links in webpage *k*. The damping factor *d* is the probability that a random surfer will follow one of the links on the present page to go to another webpage. On the other hand, 1- *d* is the probability that the surfer will open up a new page independent of the links on the existing page. The $j^{th}$ component $x_j(n)$ of *x*(*n*) is the probability that the random surfer is at the $j^{th}$ node at time *n*. The $j^{th}$ component $b_j$ of *b* is the probability that the random surfer will choose the $j^{th}$ webpage at random without following any links.

Because *T* is a Markov matrix, all the eigenvalues for *T* are contained in the closed unit disc. This guarantees that *dT* is stable, that is, all the eigenvalues for *dT* are contained in the open unit disc. In fact, all the eigenvalues for *dT* are contained in the closed disc $\mathcal{B}_d = \{\lambda \in \mathbb{C}: |\lambda| \leq d\}$ of radius *d*, in particular, $||(dT)^k|| \leq d^k v$. So the steady state solution $x(\infty)$ to (2) exists and is given by $x(\infty) = (I-dT)^{-1}b$. Finally, it is noted that the steady state solution $x(\infty)$ is independent of the initial condition *x*(0).

By recursively solving (1), we see that
$$x(n) = (dT)^n x(0) + \sum_{j=0}^{n-1}(dT)^j b \qquad (2)$$
Since *dT* is stable, the steady state solution to (1) is also determined by
$$x(\infty) = \lim_{n \to \infty} x(n) = \sum_{j=0}^{\infty}(dT)^j b = (I - dT)^{-1}b \qquad (3)$$
The last equality follows from the fact that if *A* is any matrix on $\mathbb{R}^v$ whose eigenvalues are inside the open unit disc, then $\sum_{j=0}^{\infty} A^j = (I - A)^{-1}$. Because $||(dT)^k|| \leq d^k v$, this series converges on the order of $d^n$. The components of $x(\infty)$ are the probability that a random surfer will end up at a specific webpage independent of the initial condition *x*(0). Google will thus choose the components of $x(\infty)$ with the largest probability and list those webpages first. Finally, it is noted that because $\{x(n)\}_0^{\infty}$ are all Markov vectors and *x*(*n*) converges to $x(\infty)$, it follows that $x(\infty)$ is also a Markov vector.

In our application, we use an author co-citation network in which the nodes are authors and edges represent the joint co-citations among two nodes. *T* is computed from the citation matrix *A* on $\mathbb{R}^v$ whose components $A_{jk}$ equal the number of times authors *j* and *k* are cited jointly in the same paper. *T* is thus the Markov matrix obtained by normalizing all the columns of *A* to sum to one. Following standard PageRank algorithms, we set the components $b_j = (1- d)/v$ for all *j*.

## 3. Related Works

A page with a high PageRank means that there are many pages pointing to it, or that a page with high PageRank is pointing to it. Intuitively, pages that are highly cited are worth browsing, and pages that are cited by the high PageRank pages are also worth reading. PageRank handles both cases recursively by propagating weights through the link structure of the web (Maslov & Redner, 2008).

The damping factor d is the probability that the random surfer will follow a link on the existing webpage (Brin & Page, 1998). The random surfer, however, still has a (1-d) chance to start a completely new page. A high damping factor means that the random surfer has a high chance of following the internal link, and a low chance of clicking a new external random page. Boldi, Santini, and Vigna (2005) provided the mathematical analysis of PageRank when the damping factor d changes, finding that contrary to popular belief, for real-world graphs, values of damping factor d close to 1 do not give a more meaningful ranking than other high damping factors.

A low damping factor means that every node has more or less the same chance (probability roughly equals to 1/N where N is number of nodes in the graph) to get clicked by the random surfer. The choice of damping factor is empirical and it is set to be 0.85 in Google PageRank, giving the speedy convergence of the power method (Brin & Page, 1998). Chen, Xie, Maslov, and Redner (2007) explain that the reason PageRank sets d to 0.85 is based on the observation that a typical web surfer follows the order of six hyperlinks (coincident to the six degrees of separation in social network analysis). It implies that roughly five-sixths of the time a random surfer follows the links on the webpage, while one-sixth of the time this random surfer will go to a completely new page.

Choosing a different d can result in different ranking results (Boldi et al., 2005). Damping factors in PageRank to retrieve the best match for the query can be different than those using PageRank to rank authors. In determining the best values of the damping factor for different applications, it may not be appropriate to choose low damping factors for retrieval purposes because most of the pages will have similar probabilities. Choosing d close to 1 (but not too close to 1) should give a "better" PageRank, but will significantly increase the computing complexity (Boldi et al., 2005). When d goes high, it slows down the convergence of the power method and places much greater emphasis on the hyperlink structure of the web, and much less on the random tendencies of surfers (Langville & Meyer, 2006). When d=0.99 and d=0.85, their corresponding PageRank can be vastly different. Pretto (2002) found that when d changes, the top section of the ranking changes only slightly, while the lower section of the ranking varies dramatically. As d becomes smaller, the influence of the actual link structure in the web is decreased and the effects of the artificial probability are increased. Since PageRank is trying to take advantage of the underlying link structure, it is more desirable to set d close to 1.

We can assume that articles and their citations form a scholarly network (graph). PageRank provides a computationally simple and effective way to evaluate the relative importance of publications beyond mere citation counts. The PageRank number and the number of citations to a paper when highly cited are approximately proportional to each other (Chen et al., 2007) and thus measure factors of similar importance. Researchers use weights to solve the issue of even distribution of PageRank scores among their outbound links. There are several ways to give weights to PageRank, which is normally called weighted PageRank. According to formula (1), weights can be applied to $T$ or $b$. The majority of current weighted PageRank literature focuses on adding weights to $T$ by using various normalization or application-based recalculation of T. Xing and Ghorbani (2004) added weights to the links based on its reference pages, differentiating inbound link weight and outbound link weight. Their simulation tests show that weighted PageRank performs better than the original PageRank in terms of returning larger number of relevant pages to a given query. Aktas, Nacar, and Menczer (2004) proposed a weighted PageRank based on user profiles so that weights are added to certain Internet domains users prefer. Yu, Li, and Liu (2004) added the temporal dimension to the PageRank formula and called it "TimedPageRank." They used citations and publications to define the TimePageRank formula by weighting each citation according to the citation date. Bollen, Rodriguez, and Van de Sompel (2007) proposed weighted PageRank for ranking journals and compared them with the ISI journal Impact Factor (IF). They identified some discrepancy in the top 10 rankings between ISI IF and the weighted PageRank, especially for computer science. The level of discrepancy between the ISI IF and the weighted PageRank across disciplines relates to variations in the characteristics of the publication and citation practices in different domains. Walker, Xie, Yan, and Maslov (2007) proposed a CiteRank algorithm by introducing two parameters: the inverse of the average citation depth and the time constant of the bias toward more recent publications at the start of searches. So the CiteRank shows the current popularity whereas that of PageRank corresponds to its "lifetime achievement awards" (similar to *h*-index). Liu, Bollen, Nelson, and Sompel (2005) defined AuthorRank, a modification of PageRank which considers link weight among the co-authorship links. Other work aiming at improving PageRank in the context of author ranking includes Sidiropoulos and Manolopoulos (2005) and Fiala, Rousselot, and Ježek (2008).

Based on the literature review, the authors did not find related researches on analyzing different damping factors for ranking authors based on co-citation networks. Most of the damping factor analyses are focused on improving the performance of information retrieval, the original intent for PageRank. These researches examine the range of damping factors from 0.85 to 0.99, which makes sense for retrieval purposes: setting d higher stresses the importance of the link structure and the importance of the hub and authoritative pages. There are studies using PageRank in bibliometric or scientometric researches, but few of them appear to focus on analyzing the impact of different values of damping factors for ranking purpose. Also, while most of the weighted PageRank researches focus on adding weights to *T* matrix in Formula (1), few of them place special attention on weighting the *b* part in Formula (1). This paper aims to fill in these gaps. It tests different values of damping factors, ranging from 0.05 to 0.95 for ranking authors in an author co-citation network. It proposes the weighted PageRank algorithm with adding weights to both *T* matrix and *b* part of the Formula (1), and further compares the results with other traditional methods from bibliometrics and social network analysis.

## 4. Methodology

Information Retrieval (IR) is selected as the testing field. Papers and their citations have been collected from Web of Science (WOS) from the 1970s to 2008. Based on the carefully selected search terms which are related to IR area (through checking the Library Congress Subject Heading and consulting several domain experts), we form our search strategies on the following terms, including their plurals or spelling variations: INFORMATION RETRIEVAL, INFORMATION STORAGE and RETRIEVAL, QUERY PROCESSING, DOCUMENT RETRIEVAL, DATA RETRIEVAL, IMAGE RETRIEVAL, TEXT RETRIEVAL, CONTENT BASED RETRIEVAL, CONTENT-BASED RETRIEVAL, DATABASE QUERY, DATABASE QUERIES, QUERY LANGUAGE, QUERY LANGUAGES, and RELEVANCE FEEDBACK. In total, we collected 15,370 papers with 341,871 citations. Citation records only contain first author, year, source, volume, and page number.

We select the top highly cited authors with more than 200 citations from the citation records, and in the end we arrive at 108 authors (there are some authors having the same number of citations). We calculate the co-citation frequency among each pair of authors, forming a 108-by-108 matrix. The initial PageRank score for each node (author) is 1/N (N represents the total number of nodes). The first round of PageRank calculation then starts to divide the initial PageRank score, according to different links this node has, and re-calculates the PageRank score for each node again until the PageRank for each node becomes stable (convergence). The equation of PageRank algorithm is denoted as

$$PR(p) = (1-d)\frac{1}{N} + d\sum_{i=1}^{k}\frac{PR(p_i)}{L(p_i)} \qquad (4)$$

Where $p_1, p_2, \ldots p_N$ are the pages (or nodes) under consideration, $PR(p_i)$ is the set of pages (or nodes) that link to $p_i$, $L(p_i)$ is the number of links on page (or node) $p_i$, and $N$ is the total number of pages (or nodes).

For the purpose of ranking authors, the citation structure (link structure) is important, but the random probability of citing papers is also essential. There is no rule that researchers have to refer to highly cited papers. The authors are free to cite the papers which they think are relevant rather than just referring to famous papers. We try to model this freedom in PageRank by selecting the appropriate damping factor.

The PageRank state equation Formula (1) consists of two parts: $dTx(n) + b$. The first part $dTx(n)$ models the linking structure (through weight distribution) of the citation network and the second part of $b$ represents the equal probability to get cited. The links coming from the high PageRank page to a node can increase the PageRank score of this node, especially when the damping factor is high. This means that the damping factor plays a key role in determining whether the linking structure is important. In other words, when d is high, the linking structure is important; when d is low, the influence of the linking structure is moderated.

PageRank thus provides an integrated algorithm to combine simple counting (the b part) and the topology of the network (the $dTx(n)$ part). The damping factor allows us to tune the algorithm based on the specific needs of the applications, depending on whether the topology of the network should be considered or not, and to which level. In the unusual case of the damping factor d being equal to zero, every node has the same PageRank score, which is 1/N (N represents the number of the total nodes in the network). Another extreme case is when d=1, the PageRank may become unstable and the convergence rate slows (Boldi et al., 2005).

To add weights to the PageRank, we not only add weights to the *T* matrix through the normalization of the column which converts the adjacent matrix into asymmetric matrix, but to the *b* part as well, so that nodes with high publications or citations will have a high probability of getting cited. We propose our weighted PageRank in formula (5)

$$PR\_W(p) = (1-d)\frac{W(p)}{\sum_{i=1}^{N} p_i} + d \sum_{i=1}^{k} \frac{PR(p_i)}{C(p_i)} \tag{5}$$

where *W(p)* is the weight matrix for the nodes in the graph. Each node can have different weights, which can be the number of publications of the node, or the number of citations of the node. It can be generalized to any other meaningful weights, e.g., h-index of the node, centrality measures of the node and so forth. In this study, we test Formula (5) with the weights as the publications of the author and the citations of the author. We used MATLAB to calculate the PageRank and weighted PageRank for these 108 authors based on the author co-citation network.

## 5. Results and Discussion

This section contains two parts. The first part looks at the PageRank with different damping factors and the second part deals with different weighted PageRank algorithms.

Damping Factors

Figure 1 shows the ranks of the top 20 highly cited authors. The X axis represents different damping factors ranging from 0.005 to 0.95 with 0.1 as the interval. The Y axis represents the ranking positions ranging from No. 1 to No. 55. We found that ranking positions are quite different when the damping factor d changes. We also test d=0.999 in our application. Actually there is no problem with the convergence and stability, as one is the only eigenvalue for T on the unit circle. As expected, when d=0.005, the ranking is very different when compared to the rest. It shows that when d is too low, each author has more or less the same PageRank scores (p), which are close to 1/108= 0.00926. This means that roughly every author has the same probability for any random citer to cite him/her. Even for this case, Salton G (p=0.0096), Abiteboul S (p=0.0094) from INRIA (France), and Robertson SE (p=0.0094) from Microsoft Research Cambridge (UK) are ranked the top three. Their PageRank scores are much higher than average (average p=0.00926), and they are the top three highly cited authors. If the damping factor d is very low and certain authors still rank high, it indicates that these authors really have a great influence in the network. As it does not make any sense to use d=0.005 to rank the other authors, we now consider the case when d varies from d=0.05 to d=0.95.

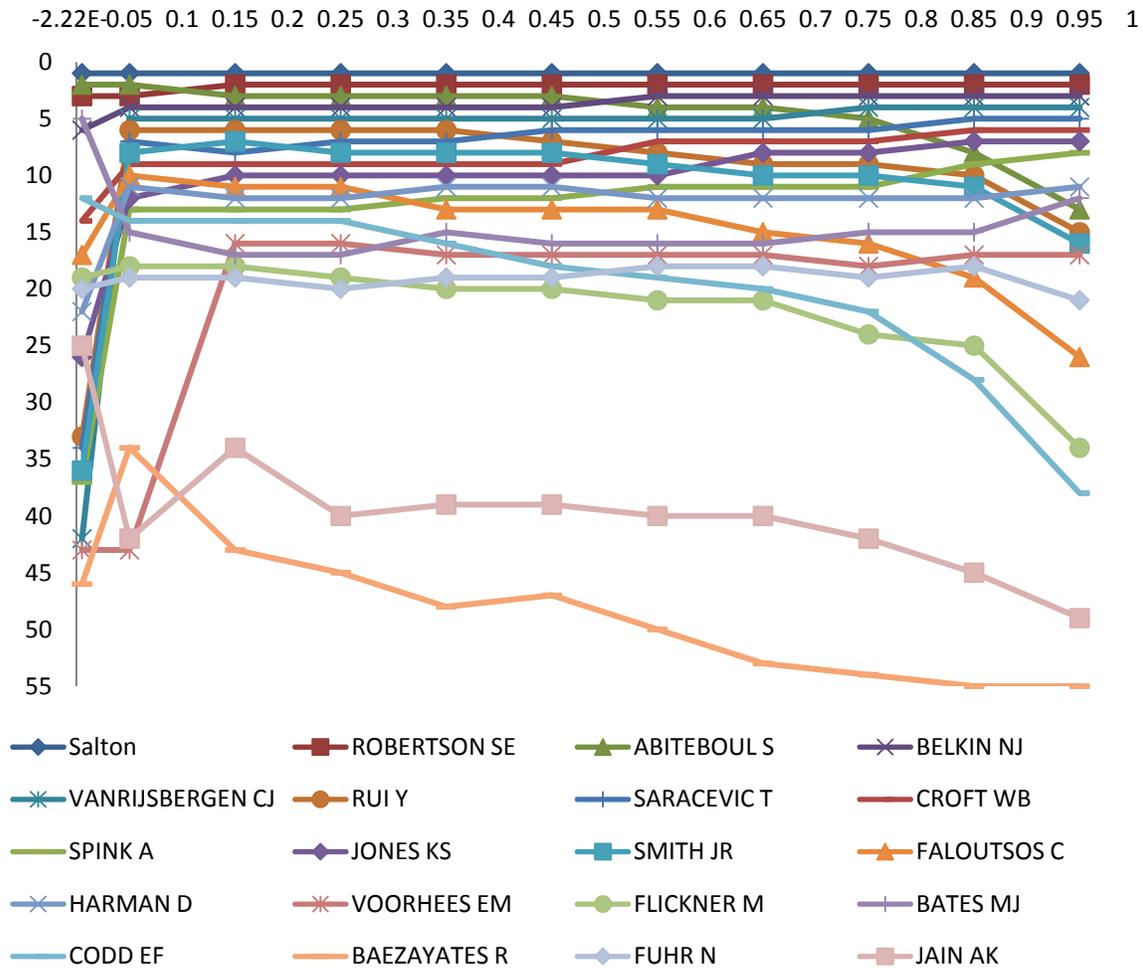

FIG. 1. Ranking of top 20 highly cited authors based on PageRank with different damping factors

Salton G and Robertson SE remain very stable for their ranking throughout different damping factors, while the ranking position of Abiteboul S drops from No. 3 (d=0.45) to No. 13 (d=0.95) when the value of damping factor is increasing. Based on the theory of PageRank algorithm, this phenomenon can be explained as follows: although Abiteboul S is co-cited with many other authors, his author co-citation frequency with Salton G is very low (only 14 times). When the value of damping factor is increasing, the citing structure becomes dominant, which means that the number of co-citations with Salton G can increase the PageRank score significantly. Abiteboul S is not co-cited frequently with Salton G, therefore his PageRank is dropping as d is increasing.

Belkin NJ and Van Rijsbergen CJ stay stable with ranking No. 3, 4 or 5 throughout different damping factors. But the rank of Rui Y is decreasing from No. 6 to No. 15, because he is only co-cited with Abiteboul S 4 times, Robertson SE 14 times, and Salton G 228 times. While he is more often co-cited with Smith JR 593 times (ranked No. 11 based on citation counting), Cox IJ 335 times (ranked No. 29 based on citation counting), and Smeulders AW 237 times (ranked No. 31 based on citation counting), with the exception of Smith JR, none of them are significant nodes in this author co-citation network. Spink A's rank is increasing from No. 13 to No. 8. Spink A is significantly co-cited together with Saracevic T (1092 times), and next with Belkin NJ (741 times), and with Jansen BJ (577 times). Her co-citations with the top three highly cited authors are Salton G (317 times), Robertson SE (101 times), and Abiteboul S (4 times). She is ranked No. 9 by citation counting, but her ranking changes according to different values of damping factor: No. 13 (d=0.05), No. 12 (d=0.35), No. 11 (d=0.55), No. 9 (d=0.85), and No. 8 (d=0.95). She is well connected to the important authors (which are referred to as the top five highly cited authors: Salton G, Robertson SE, Abiteboul S, Belkin NJ, and Van Rijsbergen CJ). When the linking structure of the network is considered heavily (when d is high), her rank will therefore go higher. The same is true for Jones KS, who based on citation is ranked No. 10. While considering the topology of the network, she is ranked No. 7. One might infer she is co-cited frequently with other important authors, and after checking the data it turns out to be true. The top three co-cited authors with her are Salton G (1447 times), Robertson SE (897 times), and Van Rijsbergen CJ (483 times).

Smith JR here is seen to drop from No. 8 (d=0.05) to No. 16 (d=0.95). Following the same analysis, we find out that the top three co-cited authors with Smith – Rui Y (593 times), Pentland A (341 times), and Flickener M (319 times) – are not "important" authors. The rank of Faloutsos C is also dropping from No. 10 to No. 26 as d is increasing. The top three co-cited authors with him are Berchtold D (391 times), Salton G (200 times), and Smith JR (170 times). Harman D stays quite stable in ranking either No. 11 or No. 12, where her top three co-cited authors are Salton G (912 times), Robertson SE (463 times), and Voorhees EM (457 times). Voorhees EM remains stable as well. Following the previous examples, we can predict that his top three co-cited authors should include the top three highly cited authors Salton G (745 times), Robertson SE (328 times), and Harman D (279 times). The rank of Flickner M is dropping significantly from No. 18 to No. 34. His co-citation status with the top three highly cited authors are Salton G (83 times), Robertson SE (5 times), and Abiteboul S (8 times), while Flickner M's top three highly co-cited authors are Smith JR (319 times), Rui Y (265 times), and Pentland A (222 times). This explains why his ranking is dropping. Bates MJ stays stable. Codd EF is changing from No. 14 to No. 38, while his citation rank is No. 17. Codd EF was co-cited with Salton G (56 times), Robertson SE (6 times), and Abiteboul S (286 times). His top three co-cited authors are Abiteboul S (286 times), Date CJ (238 times), and Ullman JD (213 times). The ranking for Baezayates R is unusual: his citation rank is No. 18 yet all his PageRank ranks are very low, ranging from No. 34 to No. 55. He is co-cited with Salton G (302 times), Robertson SE (94 times), and Abiteboul S (62 times). His top three co-cited authors are Salton G (302 times), Fuhr N (96 times), and Robertson SE (94 times). His co-citations with the rest of 107 authors are evenly distributed. The main reason for his high citation rank may be his important book *Modern Information Retrieval,* published in 1999 by Addsion-Wesley. This book is ranked No. 4 among most cited IR books. The other top three highly cited books are Salton G's *Introduction to Modern Information Retrieval* (1983), *Automatic Information Organization and Retrieval* (1968), and Van Rijsbergen CJ's *Information Retrieval* (1979). Fuhr N is also very stable, as is Jain AK. In order to find out whether different document types (mainly books or journal papers) will attract more citations in our 108 author set, we checked the top 100 highly cited documents with the authors coming from this 108 author set. We found that journal papers attract more citations (around twice as many) than books. These highly cited journal papers are not necessarily review articles. They are mainly seminal papers that offer innovations on systems or methods.

TABLE 1. Analysis of authors and their changes of PageRank

| Highly cited authors | Co-cited with the top 3 highly cited authors | | | Top 3 co-cited authors | | | PageRank |
|---|---|---|---|---|---|---|---|
| | Salton G | Robertson SE | Abiteboul S | Top1 | Top 2 | Top 3 | |
| SALTON G | 12337 | 2343 | 68 | Robertson SE (2343) | Van Rijsbergen CJ (1832) | Croft WB (1768) | Stable (always No. 1) |
| ROBERTSON SE | 2343 | 3006 | 14 | Salton G (2343) | Van Rijsbergen CJ (982) | Jones SK (897) | Stable (No. 3 or 2) |
| ABITEBOUL S | 14 | 68 | 2482 | Hull R (533) | Buneman P (487) | Ullman JD (400) | Drop (No.2 to 13) |
| BELKIN NJ | 788 | 595 | 11 | Saracevic T (1041) | Ingwersen P (828) | Salton G (788) | Stable (No. 4 or 3) |
| VANRIJSBERGEN CJ | 1832 | 982 | 12 | Salton G (1832) | Robertson SE (982) | Croft WB (634) | Stable (No. 5 or 4) |
| RUI Y | 228 | 14 | 4 | Smith JR (593) | Cox IJ (335) | Smeulders AW (237) | Drop ( No.6 to 15) |
| SARACEVIC T | 507 | 443 | 7 | Belkin NJ (1041) | Spink A (1092) | Ingwersen P (579) | Increase (No. 7 to 5) |
| CROFT WB | 1768 | 756 | 5 | Salton G (1768) | Robertson SE (756) | Jones SK (441) | Increase (No. 9 to 6) |
| SPINK A | 317 | 101 | 4 | Saracevic T (1092) | Belkin NJ (741) | Jansen BJ (577) | Increase (No. 13 to 8) |
| JONES KS | 1447 | 897 | 7 | Salton G (1447) | Robertson SE (897) | Van Rijsbergen CJ (483) | Increase (No. 12 to 7) |
| SMITH JR | 176 | 19 | 4 | Rui Y (593) | Pentland A (341) | Flickener M (319) | Drop ( No.8 to 16) |
| FALOUTSOS C | 200 | 21 | 19 | Berchtold D (391) | Salton G (200) | Smith JR (170) | Drop (No.10 to 26) |
| HARMAN D | 912 | 463 | 11 | Salton G (912) | Robertson SE (463) | Voorhees EM (457) | Stable ( No. 11 or 12) |
| VOORHEES EM | 745 | 328 | 5 | Salton G (745) | Robertson SE (328) | Harman D (279) | Increase (No. 43 to 17) |
| FLICKNER M | 83 | 5 | 8 | Smith JR (319) | Rui Y (265) | Pentland A (222) | Drop (No.18 to 34) |
| BATES MJ | 342 | 248 | 0 | Belkin NJ (725) | Saracevic T (489) | Fidel R (434) | Stable ( No. 15 or 16) |
| CODD EF | 56 | 6 | 286 | Abiteboul S (286) | Date CJ (238) | Ullman JD (213) | Drop (No.14 to 38) |
| BAEZAYATES R | 302 | 94 | 62 | Salton G (302) | Fuhr N (96) | Robertson SE (94) | Drop (No.34 to 55) |
| FUHR N | 653 | 472 | 98 | Salton G (653) | Robertson SE (472) | Van Rijsbergen CJ (362) | Stable (No. 18 or 19) |
| JAIN AK | 133 | 5 | 2 | Rui Y (210) | Smith JR (153) | Swain MJ (145) | Drop (No.42 to 49) |

Notes: Author co-citation frequencies are different when compared to author's citation frequencies. The co-citation has multiplies the two authors appearance in one paper, which leads to higher number of frequency than the authors' singular citation counts.

Table 1 summarizes the changes of the authors and their PageRank score according to different damping factors. The ranking positions of Salton G do not vary because his citation is significantly higher than the others. If the author's top three co-cited authors include some of the top five highly cited authors, their PageRank ranks either remain stable or increase when the damping factor is increasing (e.g., Spink A, Croft WB, and Voorhees EM). The PageRank for these authors drop, partly due to their co-citation with the top five highly cited authors, which are not significant compared with their top three co-cited authors (e.g., Abiteboul S, Rui Y, Smith JR, Faloutsos C, Flickner M, and Codd EF). There is still one exception, Baezayates R, who has Salton G and Robertson SE as his top three co-cited authors, but his rank is still dropping. This is perhaps due to his co-citation with the remaining 107 authors being evenly distributed, as his book is an important book in information retrieval. He is highly cited (ranked No. 18), but not highly co-cited with the other 107 authors.

Weighted PageRank

In Formula (5), the weighted PageRank, we first use the citations of the 108 authors as the weights. The weight matrix is thus one column matrix with 108 rows, where each row only has one value, the number of citations the corresponding author has. This weight matrix is further normalized by the sum of the column. So after the normalization, the total of the column sums up to one. Figure 2 shows the scatterplot of the weighted PageRank on citation (denoted as PR_c) of 108 authors in different damping factors ranging from 0.05 to 0.95. The X axis is the ranked top 108 highly cited authors with the number representing the rank of the author based on citations. It indicates that the weighted PageRank on citations does not vary much in relation to different damping factors (with average Spearman r=0.90, p<0.01). Also, the ranks of the authors, based on weighted PageRank on citations converge with the ranks of the authors based on citation counts (see Figure 2, more on Section 6).

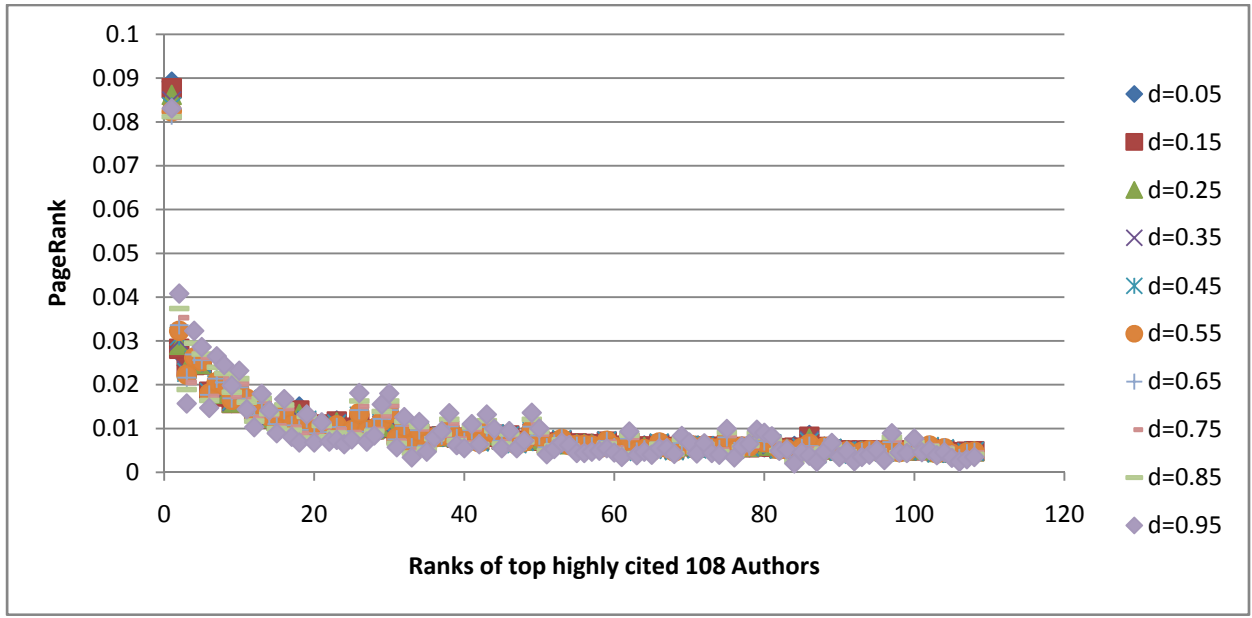

FIG 2. Scatterplot of weighted PageRank on citations of 108 authors in different damping factors

Figure 3 shows the weighted PageRank of the top 20 highly cited authors and its different damping factors. Again, we see some discrepancy between weighted PageRank ranks and the citation ranks. Notably, Abiteboul S, Rui Y, Faloutsos C, Flickner M, and Baezayates R rank lower in weighted PageRank than their corresponding citation ranks, and all of them have the corresponding drops identified in Table 1. The top three highly cited authors among these 108 authors are Salton G (average PR=0.04357 vs. average PR_c=0.08406), Robertson SE (average PR=0.02239 vs. average PR_c=0.03251), and Abiteboul S (average PR=0.01602 vs. average PR_c=0.02172). All of them have this significant increase, which can be partially explained by the weight matrix, as the "citation effect" has been stressed twice – once in the $b$ part and once in the $T$ part of Formula (1).

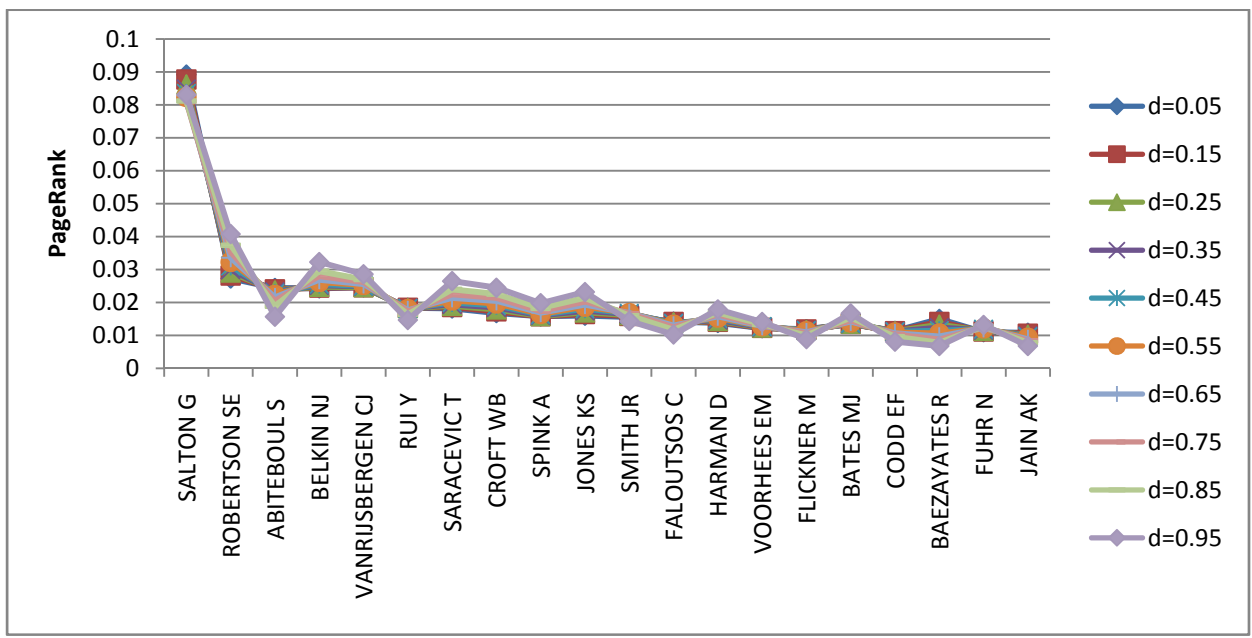

FIG 3. The top 20 highly cited authors' weighted PageRank on citations and its different damping factors

We test another weighted PageRank using the publications of the 108 authors (we only count the publications in which the 108 authors are the first authors) as the weights. The weight matrix is thus one column matrix with 108 rows, and each row only has one value, the number of publications in which the corresponding author is the first author. This weight matrix is further normalized by the sum of the column. So after the normalization, the total of the column sums up to one. Figure 4 shows the scatterplot of the weighted PageRank of 108 authors in different damping factors ranging from 0.05 to 0.95. The X axis is the top ranked 108 highly cited authors, with the number representing the rank of the author based on citation. It indicates that the weighted PageRank on publications (denoted as PR_p) does not vary much according to different damping factors (with average Spearman r=0.9 at the significant level of 0.000001). But the ranks of the authors based on weighted PageRank on publications are not consistent with the ranks of the authors based on citation counts (More on Section 6).

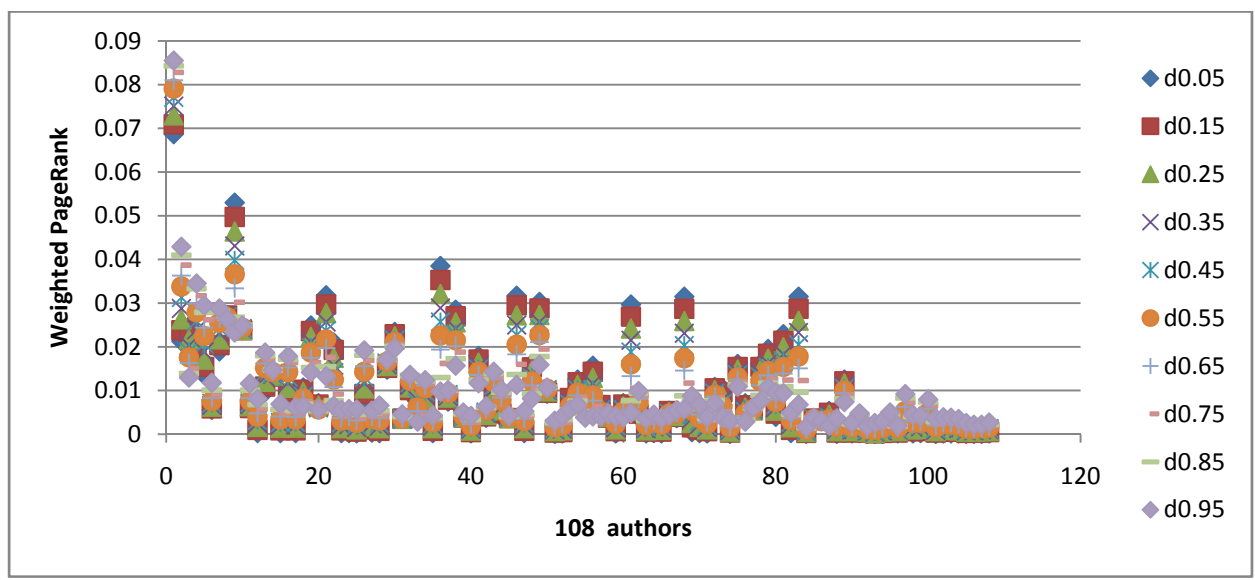

FIG 4. Scatterplot of weighted PageRank on publications of 108 authors in different damping factors

Figure 5 shows the weighted PageRank on publication of the top 20 highly cited authors and its different damping factors. Again, we see some discrepancy between weighted PageRank on publication ranks and the citation ranks. Notably, Abiteboul S, Rui Y, Faloutsos C, Flickner M, and Codd EF rank lower in the weighted PageRank on publication than their corresponding citation ranks, and all of them have the corresponding drops identified in Table 1. While Belkin NJ, Spink A, Harman D, Bates MJ, and Fuhr N's weighted PageRank on publication ranks are higher than their corresponding citation ranks, all of them either increased or stay stable in Table 1. Based on the publication ranks, the top three highly productive authors are Salton G (average PR=0.04357 vs. average PR_p=0.07775), Spink A (average PR=0.01325 vs. average PR_p=0.03823), and Chen HC (average PR=0.00843 vs. average PR_p=0.02415). All of them have a significance increase when comparing their normal PageRank with weighted PageRank on publication.

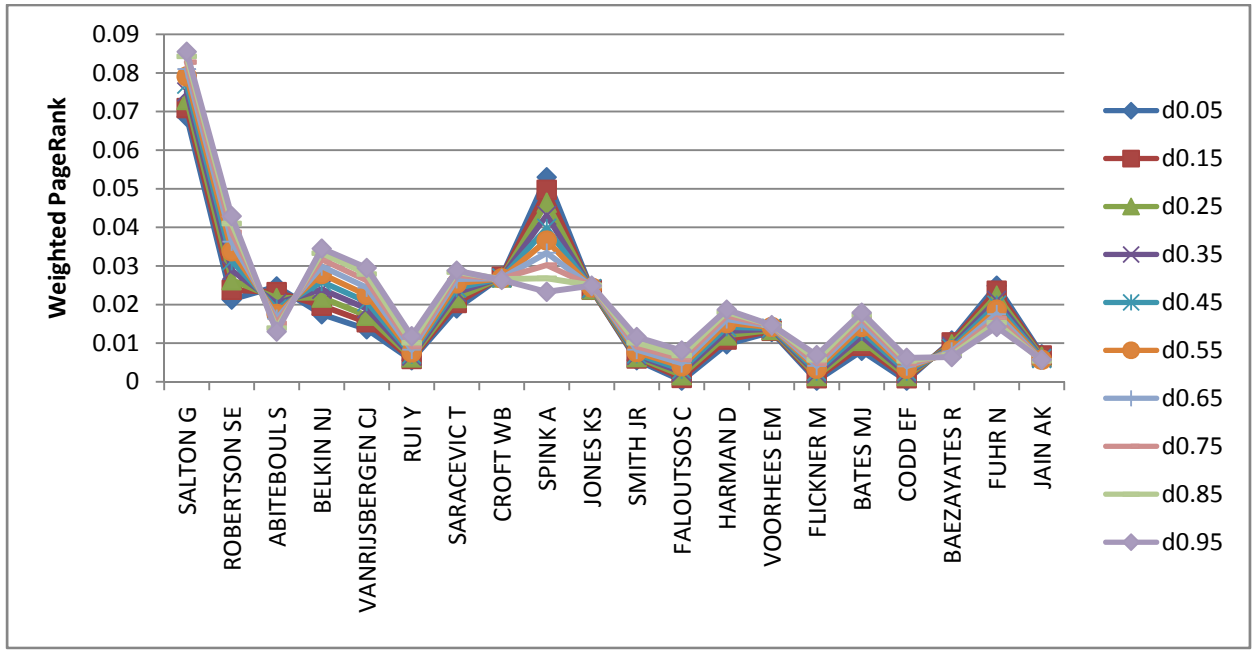

FIG 5. The top 20 highly cited authors' weighted PageRank on publication and its different damping factors

## 6. Comparison with other measures

In social network analysis, there are already well-developed methods to measure the importance of the nodes in the network (Freeman, 1979). Several researches rank authors via centrality values and compare them with scientometric ones (Liu et al., 2005; Yin, Kretschmer, Hanneman, & Liu, 2006). The centrality measures the position of each vertex in the network, and they directly associate with theories in social sciences as weak ties and social capital. They contain three major measures: degree centrality, betweenness centrality, and closeness centrality.

Degree centrality equals to the number of ties (connections) that a vertex has with other vertices. The equation where $d(n_i)$ is the degree of $n_i$ is

$$C_D(n_i) = d(n_i) \qquad (7)$$

Generally, vertices with higher degree or more connections are more central to the structure and tend to have a greater capacity to influence others. PageRank is a kind of degree centrality (directed weighted degree centrality) with the damping factor to emphasize the importance of the link structure of the graph.

A more sophisticated centrality measure is closeness (Freeman, 1979) which emphasizes the distance of a vertex to all others in the network by focusing on the geodesic distance from each vertex to all others. Closeness can be regarded as a measure of how long it will take information to spread from a given vertex to others in the network (Yin et al., 2006). Closeness centrality focuses on the extensivity of influence over the entire network. In the following equation, $C_c(n_i)$ is the closeness centrality, and $d(n_i, n_j)$ is the distance between two vertices in the network

$$C_c(n_i) = \sum_{i=1}^{N} \frac{1}{d(n_i, n_j)} \tag{8}$$

Betweenness centrality is based on the number of shortest paths passing through a vertex. Vertices with a high betweenness play the role of connecting different groups. In the following formula, $g_{jk}$ is the geodesic distance between the vertices of j and k

$$C_B(n_i) = \frac{\sum_{j<k}(n_i)}{g_{jk}} \tag{9}$$

Degree centrality identifies vertices that are locally influential. Closeness centrality focuses on the extensivity of influence over the entire network. Betweenness centrality is based on the number of shortest paths passing through a vertex. In social networks, vertices with high betweenness are brokers and connectors who bring others together (Yin et al., 2006). Being between means that a vertex has the ability to control the flow of knowledge between most others. Individuals with high betweenness are pivotal in network knowledge flow, and in social science are called structural holes. According to Burt (2002), structural holes are an opportunity to broker the flow of information between people, and control the projects that bring together people from opposites sides of the hole.

TABLE 2. Different Ranks based on different methods

| Rank | 0.05 | 0.15 | 0.25 | 0.35 | 0.45 | 0.55 | 0.65 | 0.75 | 0.85 | 0.95 | Citation | Degree | Betweenness | Closeness |
|---|---|---|---|---|---|---|---|---|---|---|---|---|---|---|
| SALTON G | 1 | 1 | 1 | 1 | 1 | 1 | 1 | 1 | 1 | 1 | 1 | 1 | 1 | 1 |
| ROBERTSON SE | 3 | 2 | 2 | 2 | 2 | 2 | 2 | 2 | 2 | 2 | 2 | 11 | 21 | 10 |
| ABITEBOUL S | 2 | 3 | 3 | 3 | 3 | 4 | 4 | 5 | 8 | 13 | 3 | 4 | 7 | 71 |
| BELKIN NJ | 4 | 4 | 4 | 4 | 4 | 3 | 3 | 3 | 3 | 3 | 4 | 22 | 24 | 20 |
| VANRIJSBERGEN CJ | 5 | 5 | 5 | 5 | 5 | 5 | 5 | 4 | 4 | 4 | 5 | 5 | 13 | 4 |
| RUI Y | 6 | 6 | 6 | 6 | 7 | 8 | 9 | 9 | 10 | 15 | 6 | 24 | 31 | 23 |
| SARACEVIC T | 7 | 8 | 7 | 7 | 6 | 6 | 6 | 6 | 5 | 5 | 7 | 25 | 51 | 25 |
| CROFT WB | 9 | 9 | 9 | 9 | 9 | 7 | 7 | 7 | 6 | 6 | 8 | 15 | 14 | 14 |
| SPINK A | 13 | 13 | 13 | 12 | 12 | 11 | 11 | 11 | 9 | 8 | 9 | 57 | 93 | 57 |
| JONES KS | 12 | 10 | 10 | 10 | 10 | 10 | 8 | 8 | 7 | 7 | 10 | 17 | 30 | 18 |
| SMITH JR | 8 | 7 | 8 | 8 | 8 | 9 | 10 | 10 | 11 | 16 | 11 | 38 | 32 | 35 |
| FALOUTSOS C | 10 | 11 | 11 | 13 | 13 | 13 | 15 | 16 | 19 | 26 | 12 | 12 | 2 | 11 |
| HARMAN D | 11 | 12 | 12 | 11 | 11 | 12 | 12 | 12 | 12 | 11 | 13 | 3 | 14 | 3 |
| VOORHEES EM | 43 | 16 | 16 | 17 | 17 | 17 | 17 | 18 | 17 | 17 | 14 | 20 | 38 | 19 |
| FLICKNER M | 18 | 18 | 19 | 20 | 20 | 21 | 21 | 24 | 25 | 34 | 15 | 16 | 11 | 15 |
| BATES MJ | 15 | 17 | 17 | 15 | 16 | 16 | 16 | 15 | 15 | 12 | 16 | 78 | 81 | 78 |
| CODD EF | 14 | 14 | 14 | 16 | 18 | 19 | 20 | 22 | 28 | 38 | 17 | 74 | 23 | 74 |
| BAEZAYATES R | 34 | 43 | 45 | 48 | 47 | 50 | 53 | 54 | 55 | 55 | 18 | 6 | 16 | 6 |
| FUHR N | 19 | 19 | 20 | 19 | 19 | 18 | 18 | 19 | 18 | 21 | 19 | 2 | 9 | 2 |
| JAIN AK | 42 | 34 | 40 | 39 | 39 | 40 | 40 | 42 | 45 | 49 | 20 | 39 | 33 | 38 |

We selected the top 20 highly cited authors and calculate their corresponding ranks by using PageRank with different damping factors, their citation ranks, and their ranks on three centrality measures (see Table 2). We also calculate the Spearman correlation among these rankings, with results shown in Table 3.

TABLE 3. The Spearman correlation analysis of different methods

| | d=0.05 | d=0.15 | d=0.25 | d=0.35 | d=0.45 | d=0.55 | d=0.65 | d=0.75 | d=0.85 | d=0.95 | Citation | Degree | Betweenness | Closeness |
|---|---|---|---|---|---|---|---|---|---|---|---|---|---|---|
| d=0.05 | 1.000 | 0.968 | 0.970 | 0.968 | 0.958 | 0.941 | 0.926 | 0.914 | 0.835 | 0.749 | 0.931 | 0.316' | 0.329' | 0.132' |
| d=0.15 | 0.968 | 1.000 | 0.998 | 0.988 | 0.983 | 0.967 | 0.956 | 0.946 | 0.878 | 0.785 | 0.958 | 0.269' | 0.240' | 0.113' |
| d=0.25 | 0.970 | 0.998 | 1.000 | 0.989 | 0.986 | 0.971 | 0.962 | 0.952 | 0.887 | 0.797 | 0.964 | 0.271' | 0.235' | 0.114' |
| d=0.35 | 0.968 | 0.988 | 0.989 | 1.000 | 0.997 | 0.985 | 0.976 | 0.971 | 0.916 | 0.841 | 0.959 | 0.262' | 0.187' | 0.101' |
| d=0.45 | 0.958 | 0.983 | 0.986 | 0.997 | 1.000 | 0.991 | 0.983 | 0.979 | 0.929 | 0.857 | 0.962 | 0.271' | 0.171' | 0.111' |
| d=0.55 | 0.941 | 0.967 | 0.971 | 0.985 | 0.991 | 1.000 | 0.995 | 0.992 | 0.959 | 0.895 | 0.965 | 0.278' | 0.171' | 0.143' |
| d=0.65 | 0.926 | 0.956 | 0.962 | 0.976 | 0.983 | 0.995 | 1.000 | 0.997 | 0.970 | 0.911 | 0.961 | 0.290' | 0.175' | 0.155' |
| d=0.75 | 0.914 | 0.946 | 0.952 | 0.971 | 0.979 | 0.992 | 0.997 | 1.000 | 0.980 | 0.931 | 0.952 | 0.271' | 0.145' | 0.156' |
| d=0.85 | 0.835 | 0.878 | 0.887 | 0.916 | 0.929 | 0.959 | 0.970 | 0.980 | 1.000 | 0.973 | 0.916 | 0.244' | 0.068' | 0.197' |
| d=0.95 | 0.749 | 0.785 | 0.797 | 0.841 | 0.857 | 0.895 | 0.911 | 0.931 | 0.973 | 1.000 | 0.836 | 0.221' | 0.019' | 0.233' |

| | | | | | | | | | | | | | |
|---|---|---|---|---|---|---|---|---|---|---|---|---|---|
| Citation | 0.931 | 0.958 | 0.964 | 0.959 | 0.962 | 0.965 | 0.961 | 0.952 | 0.916 | 0.836 | 1.000 | 0.286' | 0.198' | 0.137' |
| Degree | 0.316' | 0.269' | 0.271' | 0.262' | 0.271' | 0.278' | 0.290' | 0.271' | 0.244' | 0.221' | 0.286' | 1.000 | 0.836 | 0.842 |
| Betweenness | 0.329' | 0.240' | 0.235' | 0.187' | 0.171' | 0.171' | 0.175' | 0.145' | 0.068' | 0.019' | 0.198' | 0.836 | 1.000 | 0.652 |
| Closeness | 0.132' | 0.113' | 0.114' | 0.101' | 0.111' | 0.143' | 0.155' | 0.156' | 0.197' | 0.233' | 0.137' | 0.842 | 0.652 | 1.000 |

' is not significantly correlated; For the rest, they are significantly correlated at 0.01 level (1-tailed)

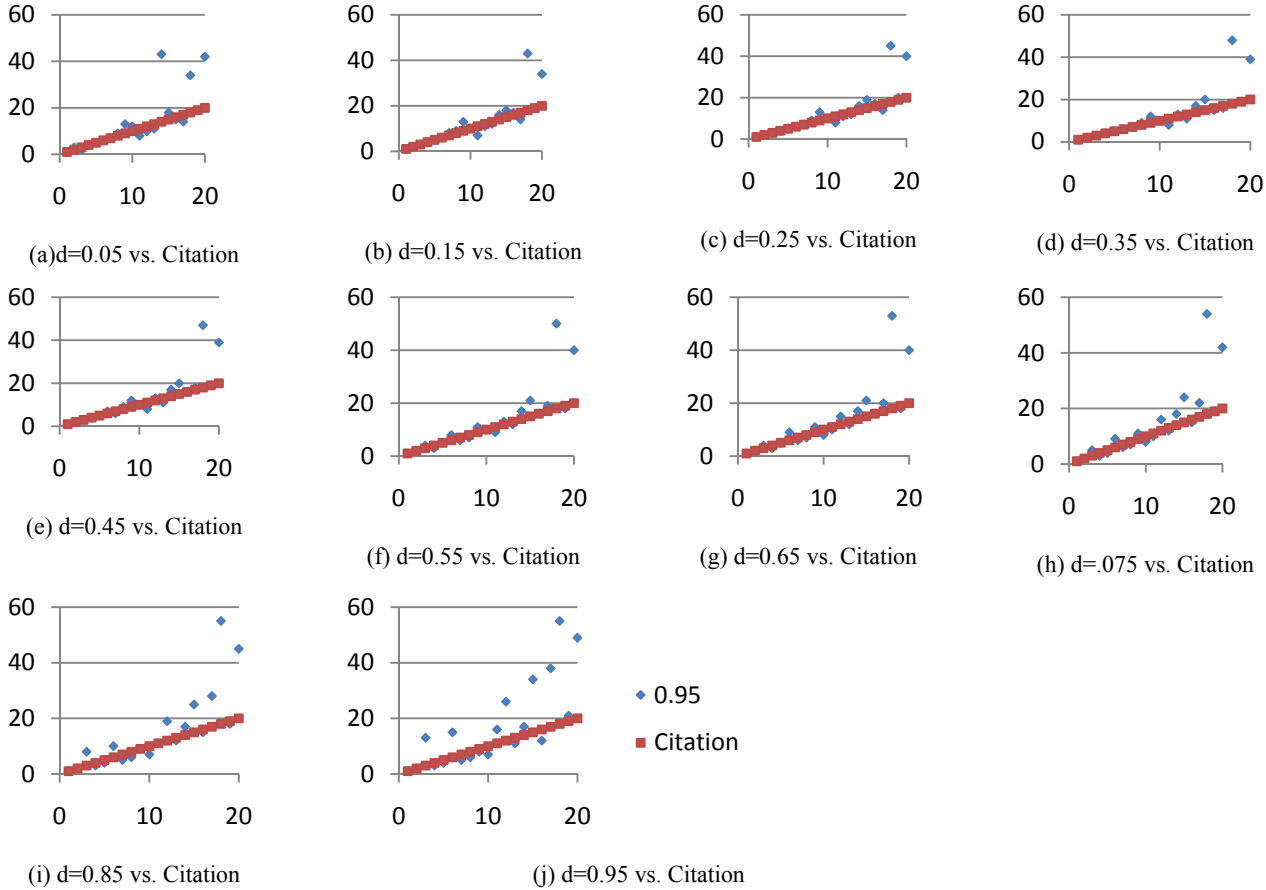

FIG 6. Scatterplots of the ranks of different damping factors versus citation ranks for the top 20 highly cited authors (with rectangle as citation rank and diamond as damping factor. X axis represents the authors which are numbered based on the ranks of their citations, and Y axis represents ranking.)

According to the Spearman correlation analysis, citation rank is highly correlated with PageRank in relation to different damping factors (average Spearman r=0.941, p<0.01). Figure 6 shows the scatterplot of the rank of PageRank with the damping factor from 0.05 to 0.95 versus citation rank. It confirms the high correlation between citation rank and different damping factors (only d=0.05, 0.85 and 0.95 are slighted less highly correlated). In Figure 6, BAEZAYATES R and JAIN AK show some extreme ranks, as most of the time they are ranked among 40s or 50s. Figure 6 is consistent with the result found by Pretto (2002) that when d changes, the top ranks stay stable while the low ranks vary. To a slight degree, the citation rank is closest to PageRank with damping factor d=0.55, which is consistent with the result of Chen, Xie, Maslov, & Redner (2007). Their empirical study shows that scientific papers usually follow a shorter path of on average two links, while the usual individual surfer follows an order of six hyperlinks (the original PageRank set $d = 1 - 1/6 \approx 0.85$). They therefore set up their damping factor as $d = 1 - \frac{1}{2} = 0.5$. Ma, Guan, and Zhao (2008) also set their damping factor to 0.5, using PageRank algorithm to evaluate the research influence for several countries in the field of Biochemistry and Molecular Biology during the period of 2000 to 2005. Yet both PageRank and citation rank are not significantly correlated with centrality measures (average Spearman r=0.197, p>0.05). In summary, in the author co-citation graph, through the analysis of the top 20 highly cited authors is based on different damping factors, we found that
- Citation rank is highly correlated with PageRank with different damping factors; and
- Citation rank and PageRank are not significantly correlated with centrality measures.

In order to further test the correlation among these measures, we extend our test to include all the 108 authors. We take normal PageRank (denoted as PR) with damping factor as 0.15 (to emphasize the equal chance to get cited), 0.55 (the short path in citation graph is around two links), and 0.85 (to stress the graph topology of citation graph); weighted PageRank on citations (PR_c) with damping factor as 0.15, 0.55 and 0.85; weighted PageRank on publications (PR_p) with damping factor as 0.15, 0.55 and 0.85; and centrality measures (degree, betweenness, closeness), h-index and citation. We compare their ranking similarity by using Spearman correlation coefficient (see Table 4). Citation rank is highly correlated with weighted PageRank on citations, then with normal PageRank, and last with weighted PageRank on publications. It has low correlation with centrality measures and h-index (but they are still significantly correlated (average Spearman r=0.403, P<0.01)). H-index is not highly correlated with the rest of the measures, making it a unique measure for ranking authors. Centrality measures have low correlation with normal PageRank and weighted PageRank (except that betweenness is not significantly correlated with weighted PageRank on publications). This indicates that centrality measures can measure different perspectives of the author impact than other measures. Normal PageRank and weighted PageRank on citations are highly correlated, while weighted PageRank on publications has relatively low correlation with the other two PageRank measures.

TABLE 4. The Spearman correlation analysis of different methods

| | PR(.15) | PR(.55) | PR(.85) | PR_c(.15) | PR_c(.55) | PR_c(.85) | PR_p(.15) | PR_p(.55) | PR_p(.85) | Degree | Betweenness | Closeness | h-index | Citation |
|---|---|---|---|---|---|---|---|---|---|---|---|---|---|---|
| PR(.15) | 1.000 | 0.971 | 0.904 | 0.883 | 0.926 | 0.880 | 0.476 | 0.558 | 0.647 | 0.288 | 0.303 | 0.258 | 0.182* | 0.819 |
| PR(.55) | 0.971 | 1.000 | 0.971 | 0.874 | 0.958 | 0.950 | 0.526 | 0.627 | 0.737 | 0.276 | 0.225 | 0.247 | 0.159' | 0.791 |
| PR(.85) | 0.904 | 0.971 | 1.000 | 0.847 | 0.957 | 0.992 | 0.597 | 0.713 | 0.846 | 0.330 | 0.171* | 0.305 | 0.146' | 0.757 |
| PR_c(.15) | 0.883 | 0.874 | 0.847 | 1.000 | 0.955 | 0.868 | 0.554 | 0.629 | 0.706 | 0.452 | 0.384 | 0.426 | 0.240 | 0.982 |
| PR_c(.55) | 0.926 | 0.958 | 0.957 | 0.955 | 1.000 | 0.966 | 0.588 | 0.686 | 0.794 | 0.397 | 0.285 | 0.371 | 0.211* | 0.895 |
| PR_c(.85) | 0.880 | 0.950 | 0.992 | 0.868 | 0.966 | 1.000 | 0.627 | 0.742 | 0.876 | 0.389 | 0.188* | 0.365 | 0.160* | 0.784 |
| PR_p(.15) | 0.476 | 0.526 | 0.597 | 0.554 | 0.588 | 0.627 | 1.000 | 0.975 | 0.857 | 0.240 | -0.039' | 0.220* | 0.242 | 0.525 |
| PR_p(.55) | 0.558 | 0.627 | 0.713 | 0.629 | 0.686 | 0.742 | 0.975 | 1.000 | 0.941 | 0.296 | -0.023' | 0.278 | 0.227 | 0.586 |
| PR_p(.85) | 0.647 | 0.737 | 0.846 | 0.706 | 0.794 | 0.876 | 0.857 | 0.941 | 1.000 | 0.393 | 0.035' | 0.378 | 0.182* | 0.644 |
| Degree | 0.288 | 0.276 | 0.330 | 0.452 | 0.397 | 0.389 | 0.240 | 0.296 | 0.393 | 1.000 | 0.771 | 0.980 | 0.126' | 0.472 |
| Betweenness | 0.303 | 0.225 | 0.171* | 0.384 | 0.285 | 0.188* | -0.039' | -0.023' | 0.035' | 0.771 | 1.000 | 0.745 | 0.145' | 0.420 |
| Closeness | 0.258 | 0.247 | 0.305 | 0.426 | 0.371 | 0.365 | 0.220* | 0.278 | 0.378 | 0.980 | 0.745 | 1.000 | 0.121' | 0.446 |
| h-index | 0.182* | 0.159' | 0.146' | 0.240 | 0.211* | 0.160* | 0.242 | 0.227 | 0.182* | 0.126' | 0.145' | 0.121' | 1.000 | 0.272 |
| Citation | 0.819 | 0.791 | 0.757 | 0.982 | 0.895 | 0.784 | 0.525 | 0.586 | 0.644 | 0.472 | 0.420 | 0.446 | 0.272 | 1.000 |

*correlation is significant at the 0.05 level (1-tailed); ' correlations is not significant; For the rest, correlation is significant at the 0.01 level (1-tailed)

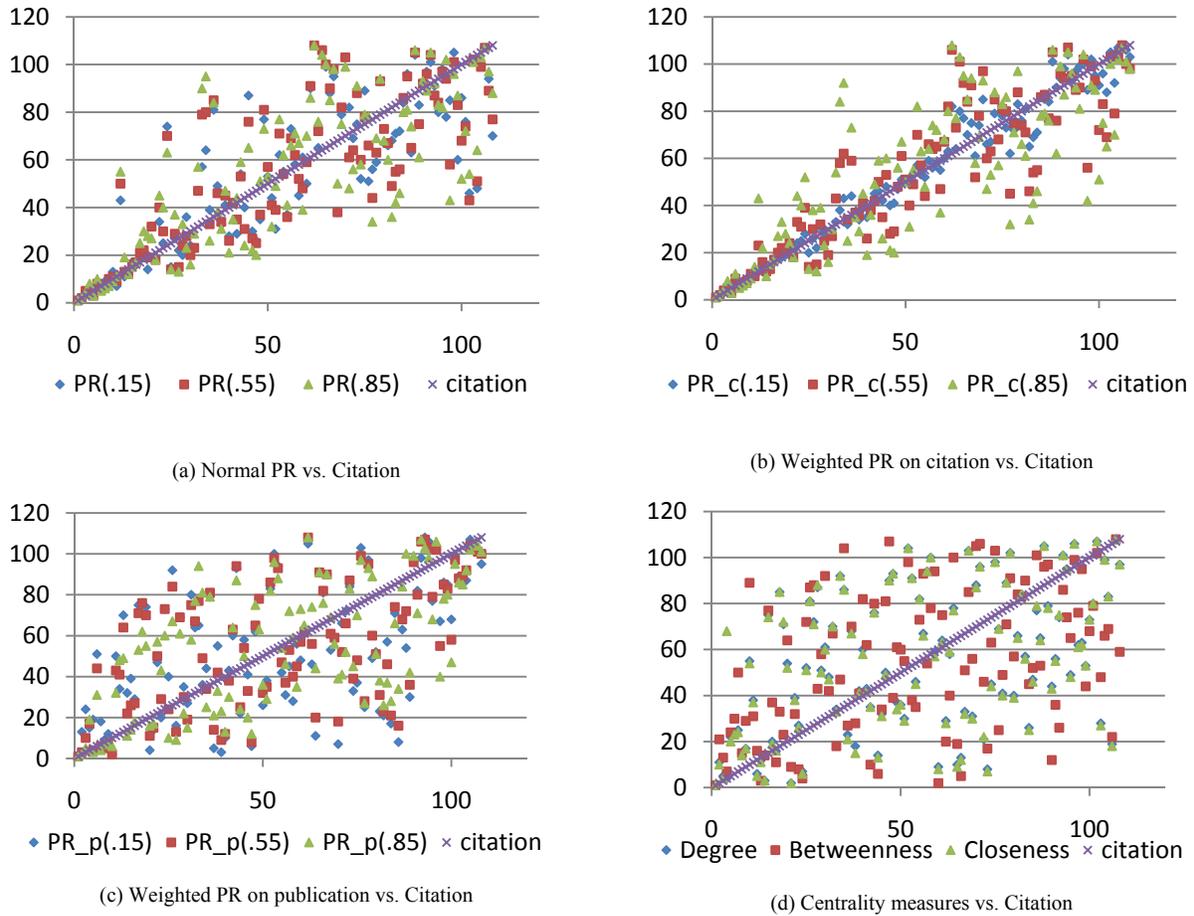

FIG 7. Scatterplot of different weighted PageRank and centrality measures versus citation rank based on the author co-citation network with the total 108 authors (X axis represents the authors which are numbered based on their citation ranks. Y axis represents the ranks.)

Figure 7 shows the scatterplots of different measures on the author co-citation network with the total 108 authors. (a) shows the normal PageRank versus citation rank, which reflects the high correlation among them, where the consistency appears in the top ranks and then slowly becomes diverse in the low ranks. (b) plots better correlation between weighted PageRank on citations versus citation ranks. (c) displays the lesser convergence between weighted PageRank on publications versus citation ranks (average Spearman r=0.585 and p<0.01). (d) illustrates the correlation among different centrality measures with citation ranks. Although they are quite scattered, they still have significant correlation with average Spearman r=0.446 and p<0.01.

In summary, based on the author co-citation graph, with the test of these 14 different measures on the total 108 authors, we identified the following:
- Citation rank is significantly correlated with all these 14 measures, especially with various PageRank algorithms;
- Normal PageRank and different weighted PageRank are significantly correlated;
- Centrality measures have low correlation with citation and various PageRank algorithms; and
- H-index is not significantly correlated with centrality measures.

# 7. Conclusion

In this study, we selected 108 most highly cited authors in the IR area from the 1970s to 2008, and formed the author co-citation network. We calculated the ranks of these 108 authors based on PageRank with damping factor ranging from 0.05 to 0.95. Furthermore, we compared the PageRank result with the citation ranking and centrality ranking.

We found that the citation rank is close to PageRank with d=0.55 (Spearman r=0.965, p<0.01). In general, the citation rank is similar to PageRank (with average spearman r=0.941, p<0.01). This means that PageRank and citation rank share similar results. In our dataset and the author co-citation network, PageRank in a certain sense can thus represent citation rank. Both citation rank and PageRank are quite different when compared to centrality ranks (average Spearman r is below 0.197, p>0.05). This means that they are not significantly correlated and their results can be varied. Citation and PageRank results are thus unable to represent the results of centrality measures.

We also introduced two different weighted PageRank algorithms: one weighted on citations and the other weighted on publications. Weighted PageRank on citations converges with the citation rank, while weighted PageRank on publications shows some discrepancies with citation ranks. We further compared the selected 14 measures based on the total 108 author set. We found that citation rnaks are highly correlated with various PageRank algorithms; normal PageRank and weighted PageRank algorithms are correlated; and centrality measures and h-index show the different perspectives of measures when compared with citation ranks and various PageRank algorithms.

Citation networks are different from hyperlink networks. They are formed by publications and the citations which cited them, while hyperlink networks are formed by the websites and hyperlinks which linked them. The major differences are:
- Unlike hyperlinks, one paper cannot update its references after it gets published, while webpages can always update their links. Google recalculates the PageRank each time it crawls the web and rebuilds its index. So in hyperlink networks, PageRank has to handle the newly created webpages and the old webpages with new updates. In citation networks, PageRank only needs to handle the newly created papers as the already published papers cannot be updated;
- Unlike hyperlinks which can point to any pages on the web, a paper can only cite published works. So citation networks introduce a time order which makes the aging effects more important than those on the web. Webpages can always update their links, so that they can even link to other webpages which are created afterwards, while papers can only cite existing publications;
- How researchers cite is also different when compared to how random surfers surf. Although there is no stipulated rule that authors should cite highly cited papers, the actual distributions of citations are skewed in Bradford-like shapes: with a mixture of few rhetoric highly cited references (e.g., monographs or seminal papers), a substantial portion of cited works of relevance to the actual topic, and a long tail of references to previous works directly used in the citing works. Researchers also tend to cite works that they read from a journal, from a conference, or from their collaborators. This phenomenon of reference behavior does demonstrate some conventions that might need to be considered when using PageRank (Cronin, Snyder, Rosenbaum, Martinson, & Callahan, 1998).

PageRank is a form of "lifetime contribution award" (same to h-index), as there is no time feature in this algorithm. In the future, we would like to add the time dimension to the PageRank algorithm for ranking authors to better reflect the dynamics of their contributions and their changes of importance over a certain period. We are also interested in testing various PageRank algorithms using different damping factors on different scholarly networks, such as citation networks formed by publications and citations (which are similar to the current Web network), and collaboration networks based on co-authorship.

## Acknowledgement

The authors would like to thank two anonymous reviewers for their valuable comments and suggestions which help a lot to better shape and improve this paper.